\documentclass{article}%
\usepackage{amsfonts}
\usepackage{amsmath}
\usepackage{amssymb}
\usepackage{graphicx}%
\newtheorem{theorem}{Theorem}

\newtheorem{definition}[theorem]{Definition}

\newtheorem{proposition}[theorem]{Proposition}

\newenvironment{proof}[1][Proof]{\noindent\textbf{#1.} }{\ \rule{0.5em}{0.5em}}
\begin{document}

\title{Coherent states for a 2-sphere with a magnetic field}
\author{Brian C. Hall and Jeffrey J. Mitchell}
\maketitle

\begin{abstract}
We consider a particle moving on a 2-sphere in the presence of a constant
magnetic field. Building on our earlier work in the nonmagnetic case, we
construct coherent states for this system. The coherent states are labeled by
points in the associated phase space, the (co)tangent bundle of $S^{2}.$ They
are constructed as eigenvectors for certain annihilation operators and
expressed in terms of a certain heat kernel. These coherent states are
\textit{not of Perelomov type}, but rather are constructed according to the
\textquotedblleft complexifier\textquotedblright\ approach of T. Thiemann. We
describe the Segal--Bargmann representation associated to the coherent states,
which is equivalent to a resolution of the identity.

\end{abstract}

\textbf{First author's address:}

University of Notre Dame

Department of Mathematics

255 Hurley Building

Notre Dame IN 46556-4618 U.S.A.

bhall@nd.edu

Supported in part by NSF Grant DMS-1001328

\textbf{Second author's address:}

Robert Morris University

Department of Mathematics

6001 University Boulevard

Moon Township PA 15108 U.S.A.

mitchellj@rmu.edu

\section{Introduction}

In \cite{H1}, Hall introduces a unitary Segal--Bargmann transform for the
group manifold of an arbitrary compact Lie group, mapping to an $L^{2}$-space
of holomorphic functions on the associated complex group. (See also
\cite{bull} for a survey of related results, \cite{ymcoherent,Wr} for
connections to the quantization of $(1+1)$-dimensional Yang--Mills theory, and
\cite{geoquant} for connections to geometric quantization.) The transform
consists of integrating the position wavefunction against certain coherent
states, which are expressed in terms of the heat kernel on the group. In
Section 11 of \cite{H1}, this transform is extended to compact symmetric
spaces, such as the $d$-sphere $S^{d}.$ M. Stenzel \cite{St1} has given a
particularly nice description of the transform for symmetric spaces, a
description that brings out the role of the heat kernel for the dual
noncompact symmetric space. The unitarity of the Segal--Bargmann transform can
be expressed, in typical physics terminology, as a resolution of the identity
for the associated coherent states. Work has also been done on
\textit{noncompact} symmetric spaces \cite{HM3,HM4,HM5,KOS,OS,OS2}, but the
situation there is much more complicated. (See also \cite{KTX} for a
Segal--Bargmann transform for the Heisenberg group.)

In a previous paper \cite{HM1}, we considered coherent states for a particle
moving in a $d$-dimensional sphere. This means that we regard $S^{d}$ as the
\textit{configuration space} of our system, with the associated phase space
then being the cotangent bundle $T^{\ast}S^{d},$ which may be identified with
the tangent bundle $TS^{d}$. (It is possible to regard the 2-sphere $S^{2}$ as
the \textit{phase space} of a classical system, but that is a completely
different problem.) Although the results about coherent states on $S^{d}$ are
in principle special cases of results of Hall and Stenzel, we gave a
self-contained and substantially different treatment of the subject. In
particular, we brought in the \textquotedblleft complexifier\textquotedblright%
\ method of T. Thiemann \cite{T1} and the \textquotedblleft polar
decomposition\textquotedblright\ method of Kowalski and Rembieli\'{n}ski
\cite{KR1}. (The coherent states in \cite{KR1} were constructed independently,
without any knowledge of the work of Hall or Stenzel.) We gave an elementary
proof of the resolution of the identity for the coherent states on $S^{d}$
(compare \cite{KR1a} in the 2-dimensional case), showing very concretely how
the heat equation on the dual noncompact symmetric space, namely
$d$-dimensional hyperbolic space, arises. We have also shown \cite{HM2} that
when $d$ is odd, the coherent states we construct converge to the usual
Gaussian coherent states on $\mathbb{R}^{d}$ in the limit as the radius of the
sphere tends to infinity. (The same result is expected to hold in the
even-dimensional case.) In the case of $S^{3}=\mathrm{SU}(2),$ many detailed
properties of the coherent states were worked out in \cite{TW}, with
applications to quantum gravity.

In the present paper, we consider a charged particle moving in $S^{2}$ in the
presence of a magnetic field of constant magnitude $B,$ pointing in the
direction perpendicular to the sphere. (If we think of our particle as a
3-dimensional particle that is constrained to move on $S^{2},$ then the
magnetic field may be thought of as coming from a magnetic monopole at the
origin.) Since, as we will see, the quantum Hilbert space for such a particle
is not the same as in the nonmagnetic case, the coherent states will
necessarily have to be modified.

In Section \ref{complex.sec}, we use the \textquotedblleft
complexifier\textquotedblright\ method of T. Thiemann to construct a
diffeomorphism $\mathbf{a}$ between the phase space $TS^{2}$ and the complex
sphere
\[
S_{\mathbb{C}}^{2}=\left\{  \left.  \mathbf{a}\in\mathbb{C}^{3}\right\vert
a_{1}^{2}+a_{2}^{2}+a_{3}^{2}=r^{2}\right\}  .
\]
When the magnetic field strength $B$ is set equal to zero, this diffeomorphism
reduces to the one considered in \cite{HM1}. In Section \ref{annihilation.sec}%
, we then use the quantum version of the complexifier method to construct
annihilation operators $A_{k}$ satisfying $A_{1}^{2}+A_{2}^{2}+A_{3}^{2}%
=r^{2}.$ In Section \ref{coherent.sec}, we construct our coherent states as
simultaneous eigenvectors for the $A_{k}$'s. In the position representation,
the coherent states can be expressed in terms of the heat kernel for a certain
line bundle over $S^{2}.$

We then turn, in Section \ref{bargmann.sec}, to the construction of a
(Segal--)Bargmann representation for the quantum Hilbert space,which is
equivalent to a resolution of the identity. The density used in the definition
of the Segal--Bargmann space is again a sort of bundle heat kernel, which may
be constructed by the method of \textquotedblleft reduction to the group
case.\textquotedblright\ Once the Segal--Bargmann space has been constructed,
we describe a unitary Segal--Bargmann transform between the position Hilbert
space and the Segal--Bargmann space. This transform consists of applying the
bundle heat operator to a section over the real sphere $S^{2}$ and then
analytically continuing to the complex sphere $S_{\mathbb{C}}^{2}.$

When the magnetic field is zero, the complex structure we get on $TS^{2}$ by
identifying it with $S_{\mathbb{C}}^{2}$ coincides with the \textquotedblleft
adapted complex structure\textquotedblright\ on $TS^{2},$ as introduced
independently by Lempert-- Sz\H{o}ke \cite{LS,Sz1} and Guillemin--Stenzel
\cite{GStenz1,GStenz2}. Meanwhile, in \cite{HK1}, the construction of the
adapted complex structure is interpreted in terms of the \textquotedblleft
imaginary-time geodesic flow,\textquotedblright\ following the complexifier
approach of Thiemann. (\textquotedblleft Time\textquotedblright\ here should
not be understood as physical time but simply as the parameter in a flow.)
More recently, Hall and Kirwin have introduced a \textquotedblleft
magnetic\textquotedblright\ version of adapted complex structure \cite{HK2}.
In the case of a constant magnetic field on $S^{2},$ the complex structure on
$TS^{2}$ given by the method of \cite{HK2} (see Section 5 of \cite{HK2})
coincides with the complex structure obtained by identifying $TS^{2}$ with
$S_{\mathbb{C}}^{2}$ by means of the diffeormorphism $\mathbf{a}.$

Finally, we note that if we apply the complexifier method for a particle
moving in the plane in a constant magnetic field, we will obtain coherent
states that are expressible in terms of the heat kernel (i.e., imaginary-time
propogator) for the quantum Hamiltonian. Such coherent states \textit{do not
}agree with the coherent states introduced by I. Malkin and V. Man'ko in
\cite{MM}, nor do they agree with the coherent states introduced in by K.
Kowalski and J. Rembieli\'{n}ski in \cite{KR2}. In particular, the
\textquotedblleft complexifier\textquotedblright\ coherent states \textit{will
not} be stable under the time evolution of the system. On the other hand, the
complexifier coherent states \textit{will} pass over smoothly to the usual
minimum-uncertainty Gaussian coherent states as the magnetic field strength
tends to zero, something that seemingly cannot be true for any coherent states
that are temporally stable. After all, when the magnetic field strength is
zero, one should not expect temporally stable coherent states, because of the
phenomenon of the spreading of the wave packet. The complexifier coherent
states on the plane do have an associated Segal--Bargmann representation,
which is described in Section 4 of \cite{KTX}, with the parameter $\lambda$ in
\cite{KTX} is to be identified with the magnetic field strength. (See Section
4 of \cite{HK2} for an explicit connection between Section 4 of \cite{KTX} and
the complexifier method.)

\section{The classical mechanics of a particle in a magnetic field}

\subsection{The $\mathbb{R}^{n}$ case}

We wish to give a Hamiltonian description of the motion of a charged particle
in $\mathbb{R}^{n}$ in the presence of a time-independent magnetic field,
described by a skew-symmetric matrix $B_{jk}.$ Since we are dealing with a
single charged particle, we can incorporate the charge of the particle into
the definition of the magnetic field. The condition $\nabla\cdot B=0$ in
$\mathbb{R}^{3}$ becomes the condition that the 2-form $(1/2)B_{jk}%
(x)dx_{j}\wedge dx_{k}$ should be closed, or, equivalently, that%
\[
\frac{\partial B_{jk}}{\partial x_{l}}+\frac{\partial B_{kl}}{\partial x_{j}%
}+\frac{\partial B_{lj}}{\partial x_{k}}=0
\]
for all $j,k,l.$ It is desirable to formulate the theory in $\mathbb{R}^{n}$
in a way that makes no reference to the vector potential, since in the sphere
case there will be no globally defined vector potential. We consider, then,
position variables $x_{j}$ and \textquotedblleft kinetic\textquotedblright%
\ momentum variables $p_{j}$ along with a Poisson bracket defined by%
\begin{equation}
\{f,g\}_{B}=\frac{\partial f}{\partial x_{j}}\frac{\partial g}{\partial p_{j}%
}-\frac{\partial f}{\partial p_{j}}\frac{\partial g}{\partial x_{j}}%
+B_{jk}(x)\frac{\partial f}{\partial p_{j}}\frac{\partial g}{\partial p_{k}}
\label{bBracketForm}%
\end{equation}
(sum convention). In particular, the relations among our position and momentum
variables are%
\begin{align}
\{x_{j},x_{k}\}  &  =0\nonumber\\
\{x_{j},p_{k}\}  &  =\delta_{jk}\nonumber\\
\{p_{j},p_{k}\}  &  =B_{jk}(\mathbf{x}). \label{brackets}%
\end{align}

We then introduce a Hamiltonian $H$ by%
\begin{equation}
H(\mathbf{x},\mathbf{p})=\frac{p^{2}}{2m}. \label{hform}%
\end{equation}
The equations of motion are computed by using the general formula
$df/dt=\{f,H\}.$ Specializing to $f=x_{j}$ and to $f=p_{j}$ gives%
\begin{align}
\frac{dx_{j}}{dt}  &  =\frac{p_{j}}{m}\nonumber\\
\frac{dp_{j}}{dt}  &  =\frac{1}{m}B_{jk}(x)p_{k}. \label{eom1}%
\end{align}
Note that $p_{j}=m~dx_{j}/dt$; this relation accounts for the terminology
\textquotedblleft kinetic momentum.\textquotedblright\ Note also that in this
approach, the magnetic field enters only into the Poisson-bracket relations
(\ref{brackets}) and not into the Hamiltonian (\ref{hform}). In the case
$n=3,$ the skew matrix $B_{jk}$ can be encoded by a vector $\mathbf{B},$ in
which case the formula for the derivative of momentum becomes $d\mathbf{p}%
/dt=(\mathbf{p}/m)\times\mathbf{B}.$ (Recall that we are absorbing the charge
of the particle into the definition of the magnetic field.)

Although the approach we have just described is the best one for generalizing
to manifolds, in the $\mathbb{R}^{n}$ case, we may alternatively consider
\textquotedblleft canonical\textquotedblright\ momentum variables $\tilde
{p}_{j}$ satisfying the usual Poisson bracket relations, that is,
$\{x_{j,}\tilde{p}_{k}\}=\delta_{jk}$ and all other brackets are zero. The
Hamiltonian is then $(\tilde{p}-A)^{2}/(2m),$ where $A $ is the vector
potential for $B.$ The two types of momentum variables are related by
$p_{j}=\tilde{p}_{j}-A_{j}.$

\subsection{The manifold case}

Let $M$ be a Riemannian manifold with metric $g,$ thought of as the
configuration space for our system. The phase space is then the cotangent
bundle $T^{\ast}M.$ On $T^{\ast}M$ we have the canonical 1-form $\theta,$
which is given in local coordinates as $\theta=p_{j}dx_{j},$ along with the
canonical 2-form $\omega:=-d\theta,$ given in coordinates as $\omega
=dx_{j}\wedge dp_{j}.$ We assume $M$ is equipped with a \textquotedblleft
magnetic field,\textquotedblright\ which we model as a closed 2-form $B.$ If
$\pi:T^{\ast}M\rightarrow M$ is the projection onto the base, then the
pulled-back form $\pi^{\ast}(B)$ is a closed 2-form on $T^{\ast}M.$ In local
coordinates, we have $B=(1/2)B_{jk}(x)dx_{j}\wedge dx_{k}$ for a unique
skew-symmetric matrix $B_{jk},$ in which case $\pi^{\ast}(B)$ is given by the
same formula, but with the $x_{j}$'s now viewed as functions on $T^{\ast}M.$

We now consider the modified symplectic form $\omega^{B}$ given by $\omega
^{B}=\omega-\pi^{\ast}(B).$ In the usual sort of cotangent bundle coordinates
$\{x_{j},p_{j}\},$ we may represent $\omega^{B}$ by the matrix%
\[
\omega^{B}=\left(
\begin{array}
[c]{cc}%
B & I\\
-I & 0
\end{array}
\right)  .
\]
Then the Poisson bracket of any two functions $f$ and $g$ is defined by
$\{f,g\}=-(\omega^{B})^{-1}(df,dg).$ It is easily verified that the formula
for $\{f,g\}$ in coordinates is the same as in (\ref{bBracketForm}). In
particular, the momentum variables do not in general Poisson commute, but
rather satisfy $\{p_{j},p_{k}\}_{B}=B_{jk}(x).$ Thus, the $p_{j}$'s should be
thought of as the kinetic momenta.

We introduce the Hamiltonian%
\[
H(x,p)=\frac{1}{2m}g_{jk}(x)p_{j}p_{k}.
\]
The dynamics associated to the Hamiltonian $H$ and the symplectic form
$\omega^{B}$ are the dynamics of a charged particle moving on $M$ acted on by
the magnetic field $B$ (but no other forces). The equations of motion in
coordinates are%
\[
\frac{dx_{j}}{dt}=\{x_{j},H\}=\frac{g_{jk}(x)}{m}p_{k}%
\]
and%
\[
\frac{dp_{j}}{dt}=\{p_{j},H\}=-\frac{1}{2m}\frac{\partial g_{kl}}{\partial
x_{j}}p_{k}p_{l}+B_{jk}(x)\frac{g_{kl}(x)}{m}p_{l}.
\]
The expression for $dx_{j}/dt$ in terms of $p_{k}$ is the same as for a free
particle moving on $M,$ which justifies calling the $p_{j}$'s the kinetic
momenta. Meanwhile, the expression for $dp_{j}/dt$ differs from a free
particle by the addition of the term involving $B.$ As in the $\mathbb{R}^{n}$
case, none of the relevant formulas requires us to choose a vector potential
for $B.$

\section{\label{complex.sec}Complex coordinates on phase space}

We now specialize to the case in which our configuration space is the 2-sphere
$S^{2},$ consisting of points $\mathbf{x}\in\mathbb{R}^{3}$ such that
$x^{2}=r^{2},$ for some positive constant $r.$ On $S^{2},$ we consider a
magnetic field equal to a constant $B$ times the area form:%
\begin{equation}
\frac{1}{2}B\varepsilon_{jkl}\frac{x_{l}}{r}~dx_{j}\wedge dx_{k}.
\label{bform}%
\end{equation}
Our goal in this section is to introduce on $T^{\ast}S^{2}$ certain complex
valued functions $a_{j},$ $j=1,2,3,$ that will allow us to identify $T^{\ast
}S^{2}$ with the complex sphere
\[
S_{\mathbb{C}}^{2}=\left\{  \mathbf{a}\in\mathbb{C}^{3}|a_{1}^{2}+a_{2}%
^{2}+a_{3}^{2}=1\right\}  .
\]
Then, in the next section, we will quantize the functions $a_{j}$ to obtain
operators $A_{j}$, which we think of as annihilation operators. Our coherent
states will then be simultaneous eigenvectors for the annihilation operators.

\subsection{\label{angular.sec}Angular momentum}

We permanently identify the cotangent bundle $T^{\ast}S^{2}$ with the tangent
bundle $TS^{2},$ using the metric on $S^{2}.$ Thus, we consider%
\[
TS^{2}=\left\{  (\mathbf{x},\mathbf{p)}\left\vert x^{2}=r^{2},~\mathbf{x}%
\cdot\mathbf{p}=0\right.  \right\}  .
\]
The canonical 2-form $\omega$ is then given by%
\[
\omega_{(\mathbf{x},\mathbf{p})}((\mathbf{a},\mathbf{b}),(\mathbf{c}%
,\mathbf{d}))=\mathbf{a}\cdot\mathbf{d-b}\cdot\mathbf{c}%
\]
for all $(\mathbf{a},\mathbf{b})$ and $(\mathbf{c},\mathbf{d})$ in
$T_{(\mathbf{x},\mathbf{p})}(TS^{2}).$ We then subtract from $\omega$ the
pull-back $\pi^{\ast}(B)$ of the \textquotedblleft magnetic\textquotedblright%
\ 2-form $B$ in (\ref{bform}) under the projection map $\pi,$ where
$\pi((\mathbf{x},\mathbf{p}))=\mathbf{x}.$ The resulting form $\omega
^{B}:=\omega-\pi^{\ast}(B)$ is closed and nondegenerate. The vector
$\mathbf{p}$ is to be thought of as the kinetic momentum of the system and not
the canonical momentum.

It is convenient to calculate in terms of appropriately defined angular
momentum functions. Since $B$ is invariant under rotations, $\omega^{B}$ is
invariant under simultaneous rotations of $\mathbf{x}$ and $\mathbf{p}.$ Let
$E_{1}$ be the vector field denoting representing an infinitesimal rotation
around the $\mathbf{e}_{1}$-axis, in both $\mathbf{x}$ and $\mathbf{p},$ so
that
\[
E_{1}=x_{2}\frac{\partial}{\partial x_{3}}-x_{3}\frac{\partial}{\partial
x_{2}}+p_{2}\frac{\partial}{\partial p_{3}}-p_{3}\frac{\partial}{\partial
p_{2}}.
\]
We then define $E_{2}$ and $E_{3}$ by cyclic permutation of the indices in the
definition of $E_{1}.$ We then look for angular momentum functions
$J_{1},J_{2},J_{3}$ such that
\[
\omega^{B}(E_{j},\cdot)=dJ_{j}.
\]
These functions will have the property that $\{J_{j},f\}=E_{j}f,$ for
$j=1,2,3.$

Since $\omega^{B}$ is a sum of terms, one of which depends only on the
position variables, we may look for $J_{j}$ of the same form. It is
straightforward to check that
\begin{equation}
\mathbf{J}(\mathbf{x},\mathbf{p})=\mathbf{x}\times\mathbf{p}-rB\mathbf{x}.
\label{j}%
\end{equation}
The Poisson bracket relations involving $\mathbf{J}$ and $\mathbf{x}$ are:%
\begin{align}
\{x_{j},x_{k}\}  &  =0\nonumber\\
\{J_{j},x_{k}\}  &  =\varepsilon_{jkl}x_{l}\nonumber\\
\{J_{j},J_{l}\}  &  =\varepsilon_{jkl}J_{l}. \label{jxBrackets}%
\end{align}
The first of these relations is true in general for magnetic symplectic forms
(compare (\ref{bBracketForm})) and the second and third relations hold because
$\{J_{j},f\}$ is an infinitesimal rotation of $f.$ Although one can work out
the Poisson bracket relations involving the linear momentum by expressing
$\mathbf{p}$ in terms of $\mathbf{J}$ as $\mathbf{p}=\mathbf{J}\times
\mathbf{x}/r^{2},$ we will not have need for these relations in the present article.

Although the relations (\ref{jxBrackets}) are identical to what we have in the
$B=0$ case, we should keep in mind that the $\mathbf{J}$ function is not the
usual one. The \textquotedblleft magnetic\textquotedblright\ $\mathbf{J}$ is
distinguished from the ordinary one by the algebraic relation
\begin{equation}
\mathbf{J}\cdot\mathbf{x}=-r^{3}B. \label{jDotX}%
\end{equation}
Note also that%
\begin{equation}
J^{2}=r^{2}p^{2}+r^{4}B^{2}. \label{jSquared}%
\end{equation}
The angular momentum is a constant of motion for the dynamics associated to
the symplectic form $\omega^{B}$ and the Hamiltonian $H=p^{2}/(2m).$ Thus, the
particle's position $\mathbf{x}$ will always lie in the circle obtained by
intersecting $S^{2}$ with the plane $\mathbf{J}\cdot\mathbf{x}=-r^{3}B,$ where
$\mathbf{J}$ is the value of the angular momentum vector at the initial time.

\subsection{\label{classicalComplexifier.sec}The classical complexifier
method}

We now apply Thiemann's complexifier method (Section 2 of \cite{T1}), as we
did in \cite{HM1} in the nonmagnetic case. To do this, we take a constant
$\alpha$ (denoted $\omega$ in \cite{HM1}) with units of frequency, and we
define our \textit{complexifier function} by%
\[
\text{complexifier}=\frac{\text{energy}}{\alpha}=\frac{p^{2}}{2m\alpha}%
=\frac{J^{2}}{2m\alpha r^{2}}+\text{const.}%
\]
Since, as will be apparent shortly, adding a constant to the complexifier has
no effect on the calculations, we will ignore the constant in the expression
for the complexifier in terms of $J^{2}.$ Then, as in \cite{HM1}, we define
complex-valued functions $a_{j}$ on $TS^{2}$ by the formula%
\[
a_{j}=e^{i\{\cdot,\text{complexifier}\}}(x_{j})=\sum_{n=0}^{\infty}\left(
\frac{i}{2m\alpha r^{2}}\right)  ^{n}\frac{1}{n!}%
\underset{n}{\underbrace{\{\cdots\{\{x_{j},J^{2}\},J^{2}\},\cdots,J^{2}\}}}.
\]
(Note that replacing $J^{2}$ by $J^{2}$ plus a constant has no effect on the
value of $a_{j}$.) The \textquotedblleft$i$\textquotedblright\ in the exponent
in the formula for $a_{j}$ should not be understood as physical time, but
merely as a parameter in our construction. That is to say, we are still going
to consider quantum mechanics using ordinary (real) time.

Using (\ref{jxBrackets}) and the product rule, we calculate $\{x_{j},J^{2}\}$
to be $2\varepsilon_{jkl}J_{k}x_{l}.$ Thus, in vector notation,%
\[
\left\{  \mathbf{x},\frac{J^{2}}{2m\alpha r^{2}}\right\}  =\frac{1}{m\alpha
r^{2}}\mathbf{J}\times\mathbf{x}.
\]
Since each $J_{j}$ Poisson-commutes with $J^{2},$ as is easily verified, we
may treat $\mathbf{J}$ as a constant in computing subsequent commutators.
Thus,%
\begin{equation}
\mathbf{a}=\exp\left\{  \frac{i}{m\alpha r^{2}}\mathbf{J}\times\cdot\right\}
(\mathbf{x})=\sum_{n=0}^{\infty}\left(  \frac{i}{m\alpha r^{2}}\right)
^{n}\frac{1}{n!}\mathbf{J}\times(\cdots\mathbf{J}\times(\mathbf{J}%
\times\mathbf{x}))). \label{aForm1}%
\end{equation}
Again, the dependence of (\ref{aForm1}) on the magnetic field strength $B$ is
through the dependence of $\mathbf{J}$ on $B.$

\begin{theorem}
\label{aForm.thm}We have%
\[
\mathbf{a}(\mathbf{x,p})=(\cosh L)\mathbf{x}+i\frac{\sinh L}{L}\frac
{\mathbf{p}}{m\alpha}-\frac{\left(  \cosh L-1\right)  }{L^{2}}B\frac
{\mathbf{J}(\mathbf{x},\mathbf{p})}{m^{2}\alpha^{2}r}%
\]
where $\mathbf{J}(\mathbf{x},\mathbf{p})$ is given by (\ref{j}) and where $L$
is a dimensionless version of the total angular momentum given by%
\[
L=\frac{\left\vert \mathbf{J}(\mathbf{x},\mathbf{p})\right\vert }{m\alpha
r^{2}}=\frac{\sqrt{p^{2}+r^{2}B^{2}}}{m\alpha r}.
\]

\end{theorem}

When $B=0,$ the $\mathbf{J}$ terms drops out, $L$ becomes equal to $p/(m\alpha
r),$ and we obtain the expression for $\mathbf{a}(\mathbf{x},\mathbf{p})$ in
Equation 18 of \cite{HM1} (with $\alpha$ being identified with $\omega$ in
\cite{HM1}). We should mention that the $B=0$ formula was already well known
prior to \cite{HM1}, for example on p. 410 of \cite{Sz1}.

\begin{proof}
A simple computation shows that%
\begin{align*}
\mathbf{J}\times\mathbf{x}  &  =r^{2}\mathbf{p}\\
\mathbf{J}\times\mathbf{p}  &  \mathbf{=}-\frac{J^{2}}{r^{2}}\mathbf{x}%
-rB\mathbf{J.}%
\end{align*}
Since also $\mathbf{J}\times\mathbf{J}=0,$ the action of \textquotedblleft
cross product with $\mathbf{J}$\textquotedblright\ on the vectors
$\mathbf{x},$ $\mathbf{p},$ and $\mathbf{J}$ may be represented by the matrix%
\begin{equation}
\mathbf{J}\times\cdot=\left(
\begin{array}
[c]{ccc}%
0 & -\frac{J^{2}}{r^{2}} & 0\\
r^{2} & 0 & 0\\
0 & -rB & 0
\end{array}
\right)  . \label{jCr}%
\end{equation}
By (\ref{aForm1}), if we exponentiate $i/(m\alpha r^{2})$ times the matrix in
(\ref{jCr}), the first column of the resulting matrix will tell us the
coefficients of $\mathbf{a}(\mathbf{x},\mathbf{p})$ in terms of the vectors
$\mathbf{x},$ $\mathbf{p},$ and $\mathbf{J}.$ The exponentiation can be done
by hand or using a computer algebra program, with the result being the formula
in the theorem.
\end{proof}

\begin{theorem}
\label{s2c.thm}Let $S_{\mathbb{C}}^{2}$ denote the set%
\[
S_{\mathbb{C}}^{2}=\left\{  \left.  \mathbf{a}\in\mathbb{C}^{3}\right\vert
a_{1}^{2}+a_{2}^{2}+a_{3}^{2}=r^{2}\right\}  .
\]
Then the map $(\mathbf{x},\mathbf{p})\mapsto\mathbf{a}(\mathbf{x},\mathbf{p})$
is diffeomorphism of $TS^{2}$ onto $S_{\mathbb{C}}^{2}.$ Furthermore, we have%
\[
\{a_{j},a_{k}\}=0
\]
for all $j$ and $k.$
\end{theorem}

Note that there are no absolute values in the definition of $S_{\mathbb{C}%
}^{2},$ which is a 2-dimensional complex submanifold of $\mathbb{C}^{3}.$ If
$C=J^{2}/(2m\alpha r^{2})$ denotes the complexifier, then $\{\cdot,C\}$ is a
derivation, meaning that $\{fg,C\}=\{f,C\}g+f\{g,C\}.$ As a result, the
exponential of $i\{\cdot,C\}$ is multiplicative, by the usual power series
argument for exponentials. Thus,%
\[
a^{2}=\sum_{j}(e^{i\{\cdot,C\}}x_{j})^{2}=e^{i\{\cdot,C\}}\left(  \sum
_{j}x_{j}^{2}\right)  =e^{i\{\cdot,C\}}(r^{2})=r^{2}.
\]
This shows that $\mathbf{a}(\mathbf{x},\mathbf{p})$ is contained in
$S_{\mathbb{C}}^{2}$ for all $\mathbf{x}$ and $\mathbf{p}.$ That $\mathbf{a}$
is a diffeomorphism of $TS^{2}$ onto $S_{\mathbb{C}}^{2}$ is shown in Section
5 of \cite{HK2}. Meanwhile, $\{\cdot,C\}$ is also a derivation with respect to
the Poisson bracket, so that the exponential of $i\{\cdot,C\}$ preserves
brackets. Thus, since $x_{j}$ and $x_{k}$ Poisson commute, $a_{j}$ and $a_{k}
$ also Poisson commute.

\section{\label{annihilation.sec}The annihilation operators}

\subsection{\label{euclideanRep.sec}Representations of the Euclidean group}

We assume that the quantum Hilbert space carries an irreducible unitary
representation of the unique simply connected Lie group $G$ whose Lie algebra
is defined by the commutation relations in (\ref{jxBrackets}). This Lie
algebra is easily identified as the Lie algebra $e(3)$ of the Euclidean group
$\mathrm{E}(3)=\mathrm{SO}(3)\ltimes\mathbb{R}^{3}.$ To find the universal
cover of $\mathrm{E}(3),$ we first note that the universal cover of
$\mathrm{SO}(3)$ is $\mathrm{SU}(2),$ where the covering map $\Xi$ of
$\mathrm{SU}(2)$ onto $\mathrm{SO}(3)$ is two-to-one and onto. For
definiteness, let us choose $\Xi:\mathrm{SU}(2)\rightarrow\mathrm{SO}(3)$ so
that
\begin{equation}
\Xi\left(
\begin{array}
[c]{cc}%
e^{i\theta/2} & 0\\
0 & e^{-i\theta/2}%
\end{array}
\right)  =\left(
\begin{array}
[c]{rrr}%
\cos\theta & -\sin\theta & 0\\
\sin\theta & \cos\theta & 0\\
0 & 0 & 1
\end{array}
\right)  . \label{su2so3}%
\end{equation}
The universal cover $\mathrm{\tilde{E}}(3)$ of $\mathrm{E}(3)$ is then given
by%
\[
\mathrm{\tilde{E}}(3)=\mathrm{SU}(2)\ltimes\mathbb{R}^{3}.
\]
Here, $\mathrm{SU}(2)$ acts on $\mathbb{R}^{3}$ by first mapping to
$\mathrm{SO}(3)$ by the two-to-one covering map and then acting on
$\mathbb{R}^{3}$ by rotations.

The irreducible representations of $\mathrm{\tilde{E}}(3)$ are classified by
the Wigner--Mackey method. (See, for example, \cite{Fo2}.) To apply this
method, we first choose an orbit of $\mathrm{SU}(2)$ inside $\mathbb{R}^{3},$
which is a sphere of some radius $r$ that we assume is positive. (We identify
this radius with the radius of the sphere whose cotangent bundle we are
quantizing.) We then choose a point in $S^{2},$ which we take to be the north
pole $\mathbf{n}=(0,0,r).$ The \textit{little group} is then the subgroup of
$\mathrm{SU}(2)$ that maps $\mathbf{n}$ to $\mathbf{n}.$ From (\ref{su2so3}),
we can see that the little group is just the diagonal subgroup $D$ of
$\mathrm{SU}(2).$ The choice of an irreducible representation of the little
group then completes the specification of an irreducible representation of
$\mathrm{\tilde{E}}(3).$ Every irreducible representation of $D$ is one
dimensional and of the form%
\[
\left(
\begin{array}
[c]{cc}%
e^{i\theta/2} & 0\\
0 & e^{-i\theta/2}%
\end{array}
\right)  \mapsto e^{il\theta},
\]
for some integer or half-integer $l.$ In the notation of K. Kowalski and J.
Rembieli\'{n}ski \cite{KR1}, the parameter $l$ is the \textquotedblleft
twist\textquotedblright\ of the system; it is analogous to the spin of a
particle moving in $\mathbb{R}^{3}.$

We will use the standard basis $\{E_{1},E_{2},E_{3}\}$ of $\mathrm{su}(2),$
given by%
\begin{equation}
E_{1}=\frac{1}{2}\left(
\begin{array}
[c]{rr}%
0 & 1\\
-1 & 0
\end{array}
\right)  ;\quad E_{2}=\frac{1}{2}\left(
\begin{array}
[c]{rr}%
0 & i\\
i & 0
\end{array}
\right)  ;\quad E_{3}=\frac{1}{2}\left(
\begin{array}
[c]{rr}%
i & 0\\
0 & -i
\end{array}
\right)  . \label{fBasis}%
\end{equation}
These matrices satisfy $[E_{j},E_{k}]=\varepsilon_{jkl}E_{l}.$ Let
$\Sigma_{r,l}$ denote the representation of $\mathrm{\tilde{E}}(3)$
corresponding to a choice of $r,l.$ Then the associated Lie algebra
representation is described by \textquotedblleft position\textquotedblright%
\ operators $X_{1},X_{2},X_{3},$ whose joint spectrum is $S^{2},$ along with
\textquotedblleft angular momentum\textquotedblright\ operators $\hat{J}%
_{1},\hat{J}_{2},\hat{J}_{3}$ given by%
\[
\hat{J}_{j}=\left.  i\hbar\frac{d}{dt}\Sigma_{r,l}\left(  e^{tE_{j}}\right)
\right\vert _{t=0}.
\]
If $\psi$ is a (generalized) eigenvector for the position operators with
$X_{1}\psi=X_{2}\psi=0$ and $X_{3}\psi=r\psi,$ our choice of a representation
of the little group means that%
\[
\hat{J}_{3}\psi=\hbar l\psi.
\]

The position and angular momentum operators satisfy relations analogous to
(\ref{jxBrackets}):%
\begin{align}
\frac{1}{i\hbar}[X_{j},X_{k}]  &  =0\nonumber\\
\frac{1}{i\hbar}[\hat{J}_{j},X_{k}]  &  =\varepsilon_{jkl}X_{l}\nonumber\\
\frac{1}{i\hbar}[\hat{J}_{j},\hat{J}_{k}]  &  =\varepsilon_{jkl}\hat{J}_{l}.
\label{jxQ}%
\end{align}
The choice of a sphere of radius $r$ in the Wigner--Mackey method gives us the
additional algebraic relation%
\begin{equation}
\mathbf{X}\cdot\mathbf{X}=r^{2}. \label{xxQ}%
\end{equation}
Finally, the parameter $l,$ labeling the chosen representation of the little
group, determines one additional relation:%
\begin{equation}
\mathbf{\hat{J}}\cdot\mathbf{X}=r\hbar l. \label{jDotxQ}%
\end{equation}
To see that this relation is true, we can easily verify that $\mathbf{\hat{J}%
}\cdot\mathbf{X}$ commutes with each $\hat{J}_{j}$ and each $X_{j},$ which
means that this operator must act as a constant multiple of the identity in
each irreducible representation. The value of this constant can be determined
by evaluating on a (generalized) vector $\psi$ such that $X_{1}\psi=X_{2}%
\psi=0$ and $X_{3}\psi=r,$ on which we have $\mathbf{\hat{J}}\cdot
\mathbf{X}\psi=r\hat{J}_{3}\psi=r\hbar l\psi,$ by assumption.

Comparing (\ref{jDotxQ}) to (\ref{jDotX}) in the classical case, it is natural
to make the following identification, which relates the value of $l$ on the
quantum side to the value of $B$ on the classical side:%
\begin{equation}
-\frac{B}{r}=\frac{\hbar l}{r^{2}}. \label{bl1}%
\end{equation}
That is to say, if we make the identification (\ref{bl1}), then (\ref{jDotxQ})
becomes identical to the classical formula:%
\begin{equation}
\mathbf{\hat{J}}\cdot\mathbf{X}=-r^{3}B. \label{jDotxQ2}%
\end{equation}
Now, (\ref{bl1}) is equivalent to the condition%
\begin{equation}
-\frac{(4\pi r^{2})B}{2\pi\hbar}=2l, \label{bl2}%
\end{equation}
where $2l$ is a non-negative integer. Equation (\ref{bl2}) says that the area
of the sphere, with respect to the magnetic 2-form---which is $B$ times the
area form---must be an integer multiple of $2\pi\hbar.$ Since the restriction
of the canonical 2-form $\omega$ to $S^{2}\subset TS^{2}$ is zero, an
equivalent formulation of the condition is that the symplectic area of the
sphere $S^{2}\subset TS^{2}$ with respect to $\omega^{B}$ has to be an integer
multiple of $2\pi\hbar.$ This last condition is the usual integrality
condition in the theory of quantization of symplectic manifolds. (See, for
example, \cite{Wo}.)

Note that we have taken the position and \textit{angular} momentum operators
as the \textquotedblleft basic\textquotedblright\ operators of our theory. If
we wish to introduce \textit{linear} momentum operators, we must define them
in terms of the angular momentum operators. Since we have, classically,
$\mathbf{p}=\mathbf{J}\times\mathbf{x}/r^{2},$ it is reasonable to
\textit{define} the quantum version of $\mathbf{p}$ by the analogous relation:%
\begin{equation}
\mathbf{P}:=\frac{1}{r^{2}}\mathbf{\hat{J}}\times\mathbf{X.} \label{pQdef}%
\end{equation}
These linear momentum operators will come up in the computation of the
annihilations operators in the next subsection.

\subsection{\label{quantumComplexifier.sec}The quantum complexifier method}

We work in a Hilbert space constituting an irreducible representation of
$\mathrm{\tilde{E}}(3),$ with operators $\hat{J}_{j}$ and $X_{j}$ satisfying
the commutation relations (\ref{jxQ}) along with the algebraic relations
(\ref{xxQ}) and (\ref{jDotxQ}). We assume that the quantum counterpart
$\hat{H}$ to the classical energy function is equal to $\hat{J}^{2}/(2mr^{2})$
plus a constant, in which case our \textit{complexifier operator} is%
\begin{equation}
\text{complexifier}=\frac{\text{energy}}{\alpha}=\frac{\hat{J}^{2}}{2m\alpha
r^{2}}+\text{const.,} \label{hQ}%
\end{equation}
as in the classical case. Here, the parameter $\alpha$, having units of
frequency, is the same one used in Section \ref{classicalComplexifier.sec}. As
in the classical setting, the constant on the right-hand side of (\ref{hQ})
has no effect on the complexifier method, as will be evident shortly.

Following the quantum version of Thiemann's method \cite{T1}, we define
operators $A_{j}$ by formula%
\begin{equation}
A_{j}=e^{i[\cdot,\text{complexifier}]/(i\hbar)}(X_{j})=\sum_{n=0}^{\infty
}\left(  \frac{1}{2m\alpha r^{2}\hbar}\right)  ^{n}%
\underset{n}{\underbrace{[\cdots\lbrack\lbrack X_{j},\hat{J}^{2}],\hat{J}%
^{2}],\ldots\hat{J}^{2}]}}. \label{aQ}%
\end{equation}
We will interpret these operators as the \textit{annihilation operators} for
our system. By a standard identity (see, for example, Proposition 2.25 and
Exercise 2.19 in \cite{H11}), we have the alternative expression
\begin{equation}
A_{j}=\exp\left\{  -\frac{\hat{J}^{2}}{2m\alpha r^{2}\hbar}\right\}  X_{j}%
\exp\left\{  \frac{\hat{J}^{2}}{2m\alpha r^{2}\hbar}\right\}  . \label{aQ2}%
\end{equation}
For purposes of computing the coherent states, the expression (\ref{aQ2}) is
the most useful formula for the annihilation operators. In particular, from
(\ref{aQ2}), we can see that%
\begin{equation}
\frac{1}{i\hbar}[A_{j},A_{k}]=0 \label{AjCommute}%
\end{equation}
and%
\begin{equation}
A_{j}A_{j}=r^{2}. \label{AjSquared}%
\end{equation}

We now look for quantum counterparts to the expressions for $\mathbf{a}%
(\mathbf{x},\mathbf{p})$ in (\ref{aForm1}) and Theorem \ref{aForm.thm}. In
computing the commutator of $X_{j}$ with $\hat{J}^{2},$ we get products of $X
$'s and $\hat{J}$'s in both orders. If we move, say, all the $\hat{J}$'s to
the left we obtain a quantum correction as follows:%
\[
\frac{1}{i\hbar}\left[  \mathbf{X},\frac{\hat{J}^{2}}{2m\alpha r^{2}}\right]
=\frac{1}{m\alpha r^{2}}\left(  \mathbf{\hat{J}}\times\mathbf{X}%
-i\hbar\mathbf{X}\right)  ,
\]
as may easily be verified. Now, since $\hat{J}^{2}$ commutes with each
$\hat{J}_{j},$ we may treat $\mathbf{\hat{J}}$ as a constant in computing
subsequent commutators. Thus,%
\begin{equation}
\mathbf{A}=\exp\left\{  \frac{i(\mathbf{\hat{J}}\times\cdot)+\hbar}{m\alpha
r^{2}}\right\}  \left(  \mathbf{X}\right)  . \label{aQ3}%
\end{equation}
In the case $\hbar/(m\alpha r^{2})=1,$ this expression is essentially Equation
(4.7) of \cite{KR1}, which should be expected, since we have thus far used
only the commutation relations (\ref{jxQ}) and not (\ref{xxQ}) or
(\ref{jDotxQ}). Equation (\ref{aQ3}) also coincides with the $d=2$ case of
Equation (33) of \cite{HM1}. (The notation $\mathbf{JX}$ in \cite{HM1}
corresponds, in the $d=2$ case, to $\mathbf{\hat{J}}\times\mathbf{X}$ in the
notation of the current article.)

We now compute the annihilation operators \textquotedblleft
explicitly\textquotedblright\ in a form similar to the expressions for
$\mathbf{a}(\mathbf{x},\mathbf{p})$ in Theorem \ref{aForm.thm}. Recalling the
definition (\ref{pQdef}) of the linear momentum operators, a straightforward
computation gives:%
\begin{align*}
\mathbf{\hat{J}}\times\mathbf{X}  &  =r^{2}\mathbf{P}\\
\mathbf{\hat{J}}\times\mathbf{P}  &  =-\frac{\hat{J}^{2}}{r^{2}}%
\mathbf{X}+i\hbar\mathbf{P}-rB\mathbf{J.}%
\end{align*}
Thus, if we cross with $\mathbf{\hat{J}}$ repeatedly, we will obtain
expressions involving $\mathbf{\hat{J}}$ in addition to $\mathbf{X}$ and
$\mathbf{P}.$ In the quantum case, $\mathbf{\hat{J}}\times\mathbf{\hat{J}}$ is
not zero:%
\[
\mathbf{\hat{J}}\times\mathbf{\hat{J}}=i\hbar\mathbf{\hat{J}}.
\]
The action of the operation of \textquotedblleft crossing with $\mathbf{\hat
{J}}$\textquotedblright\ on the vector operators $\mathbf{X},$ $\mathbf{P},$
and $\mathbf{\hat{J}}$ can thus be encoded in the following matrix:%
\begin{equation}
\mathbf{\hat{J}}\times\cdot=\left(
\begin{array}
[c]{ccc}%
0 & -\frac{\hat{J}^{2}}{r^{2}} & 0\\
r^{2} & i\hbar & 0\\
0 & -rB & i\hbar
\end{array}
\right)  . \label{jCross}%
\end{equation}

Since entries of the matrix in (\ref{jCross}) commute, we can think of it as
an ordinary $3\times3$ matrix. We then put in this matrix in place of the
expression $\mathbf{\hat{J}}\times\cdot$ in (\ref{aQ3}) and exponentiate. The
matrix exponential can be computed explicitly by Mathematica, and the first
column of the exponential gives us the result of applying the quantum
complexifier to $\mathbf{X}.$ The matrix in (\ref{jCross}) is block-upper
triangular and the upper left $2\times2$ block is the same as the matrix in
\cite{HM1}; as a result, the upper $2\times2$ block in the exponential is the
same as the exponential of the matrix in \cite{HM1}. Thus, the expression for
$\mathbf{A}$ will be the same as the $d=2$ case of Equation (38) of
\cite{HM1}, except that there will be an extra term involving $\mathbf{\hat
{J}}.$ (Recall that what we call $\alpha$ here corresponds to $\omega$ in
\cite{HM1}.) We record the answer in the following theorem.

\begin{theorem}
\label{aQ.thm}Let us introduce the shifted, dimensionless angular momentum
operator%
\[
\hat{L}=\frac{\sqrt{\hat{J}^{2}+\hbar^{2}/4}}{m\alpha r^{2}}.
\]
Then we obtain%
\begin{align}
\mathbf{A}  &  =e^{\hbar/(2m\alpha r^{2})}\left(  \cosh\hat{L}+\frac{\hbar
}{2m\alpha r^{2}}\frac{\sinh\hat{L}}{\hat{L}}\right)  \mathbf{X}\nonumber\\
&  +ie^{\hbar/(2m\alpha r^{2})}\frac{\sinh\hat{L}}{\hat{L}}\frac{\mathbf{P}%
}{m\alpha}-\Lambda B\frac{\mathbf{\hat{J}}}{m^{2}\alpha^{2}r}, \label{aQ4}%
\end{align}
where%
\[
\Lambda=\frac{1}{\hat{L}^{2}-\left(  \hbar/(2m\alpha r^{2})\right)  ^{2}%
}\left\{  1+e^{\hbar/(2m\alpha r^{2})}\left(  \frac{\hbar}{2m\alpha r^{2}%
}\frac{\sinh\hat{L}}{\hat{L}}-\cosh\hat{L}\right)  \right\}  .
\]

\end{theorem}

When $B=0,$ the expression for $\mathbf{A}$ agrees (upon setting
$\hbar/(m\alpha r^{2})=1$) with Equation (4.16) of \cite{KR1a} . The $B=0$
case of Theorem \ref{aQ.thm} also agrees with the $d=2$ case of Equation (38)
in \cite{HM1}. On the other hand, taking the limit as $\hbar$ tends to zero in
(\ref{aQ4})---and identifying $\hat{L}$ with $L$---gives the expression for
the classical function $\mathbf{a}(\mathbf{x},\mathbf{p})$ in Theorem
\ref{aForm.thm}.

\section{\label{coherent.sec}The coherent states}

We define a state $\psi$ to be a \textbf{coherent state} is $\psi$ is a
simultaneous eigenvector for the operators $A_{j}$:%
\[
A_{j}\psi=a_{j}\psi.
\]
Since the $A_{j}$'s commute (see (\ref{AjCommute})) it is reasonable to hope
that there are many coherent states. Since also $A_{j}A_{j}=r^{2},$ we must
have%
\[
a_{1}^{2}+a_{2}^{2}+a_{3}^{2}=r^{2},
\]
meaning that the vector $\mathbf{a}:=(a_{1},a_{2},a_{3})$ must belong to
$S_{\mathbb{C}}^{2}\subset\mathbb{C}^{3}.$

\begin{theorem}
\label{cs.thm}For each $\mathbf{a}\in S_{\mathbb{C}}^{2},$ there exists a
nonzero, normalizable vector $\chi_{\mathbf{a}}$ in the quantum Hilbert space
such that%
\[
A_{j}\chi_{\mathbf{a}}=a_{j}\chi_{\mathbf{a}}.
\]
For $\mathbf{a}$ in the real sphere $S^{2},$ we may compute $\chi_{\mathbf{a}%
}$ as%
\[
\chi_{\mathbf{a}}=\exp\left\{  -\frac{\hat{J}^{2}}{2m\alpha r^{2}\hbar
}\right\}  \delta_{\mathbf{a}},
\]
where $\delta_{\mathbf{a}}$ is a (non-normalizable) vector satisfying
$X_{j}\delta_{\mathbf{a}}=a_{j}\delta_{\mathbf{a}}.$
\end{theorem}

For each $\mathbf{a}$ in the real sphere $S^{2},$ the space of (generalized)
eigenvectors $\psi$ for $\mathbf{X}$ satisfying $X_{j}\psi=a_{j}\psi$ is
one-dimensional. When $l\neq0,$ there is no way to pick a nonzero element
$\delta_{\mathbf{a}}$ of each eigenspace that depends continuously on
$\mathbf{a}.$ Thus, there is no continuous way to parameterize the coherent
states as vectors, even for parameters in the real sphere. Physically,
however, it is only the one-dimensional subspace spanned by the coherent state
that is important, and these subspaces depend continuously (in fact,
holomorphically) on $\mathbf{a}\in S_{\mathbb{C}}^{2}.$

\begin{proof}
As explained in detail in Section \ref{bargmann.sec}, there is a
\textquotedblleft position representation\textquotedblright\ in which our
Hilbert space is the space of square-integrable sections of complex line
bundle over $S^{2}.$ Then for $\mathbf{a}\in S_{\mathbb{C}}^{2}$, the coherent
state $\chi_{\mathbf{a}}$ is nothing but the \textquotedblleft bundle heat
kernel,\textquotedblright\ evaluated at a point in the fiber over
$\mathbf{a}.$ It is well known that such bundle heat kernel always exists and
is smooth, so that $\psi_{\mathbf{a}}$ is a normalizable (finite-norm) vector.
For general $\mathbf{a}\in S_{\mathbb{C}}^{2},$ we need to show that the
bundle heat kernel can be analytically continued with respect to the parameter
$\mathbf{a}$ from $S^{2}$ to $S_{\mathbb{C}}^{2}.$ It suffices to show that
any solution of the bundle heat equation can be analytically continued from
$S^{2}$ to $S_{\mathbb{C}}^{2},$ which we will show in Section
\ref{bargmann.sec} by the method of reduction to the group case. (In the group
case, the existence of the analytic continuation of the heat kernel was shown
in detail in Section 4 of \cite{H1}.)
\end{proof}

Computations of the heat kernel for the \textquotedblleft
spinor\textquotedblright\ case ($l=1/2$) can be found in \cite{Ca}.

\section{The Segal--Bargmann representation\label{bargmann.sec}}

In this section we construct a Segal--Bargmann representation associated to
the coherent states, and an associated unitary Segal--Bargmann transform. That
is to say, the transform consists of taking the inner product of a state
$\psi$ with each coherent state $\chi_{\mathbf{a}},$ resulting in a function
of $\mathbf{a}.$ Because the coherent states depend holomorphically on
$\mathbf{a},$ we get a holomorphic function on $S_{\mathbb{C}}^{2},$ or
rather, a holomorphic section of a certain line bundle over $S_{\mathbb{C}%
}^{2}.$ The unitarity of the Segal--Bargmann transform is equivalent to
resolution of the identity for the coherent states, as we explain in Section
VII of \cite{HM1}.

\subsection{The Schr\"{o}dinger Hilbert space}

In Section \ref{euclideanRep.sec}, we considered irreducible representations
of the double cover of the Euclidean group \textquotedblleft in the
abstract.\textquotedblright\ That is, we never give a concrete realization of
the Hilbert space, but rather perform all calculations using only the
commutation relations of the Lie algebra together with the two algebraic
relations (\ref{xxQ}) and (\ref{jDotxQ}) that characterize the particular
irreducible representation. We now wish to give a concrete realization of a
given irreducible representation, as a space of square-integrable
\textquotedblleft sections\textquotedblright\ over the real sphere.

If $(\Pi,V)$ is a finite-dimensional representation of $\mathrm{SU}(2),$
define operators $\sigma_{j}$ by%
\begin{equation}
\sigma_{j}=i\hbar\left.  \frac{d}{dt}\Pi\left(  e^{tE_{j}}\right)  \right\vert
_{t=0}, \label{sigmaj}%
\end{equation}
where $E_{j}$ is defined in (\ref{fBasis}). For each non-negative integer or
half-integer $l,$ let $(\Pi_{l},V_{l})$ be the irreducible unitary
representation of $\mathrm{SU}(2)$ in which the largest eigenvalue of
$\sigma_{3}$ is $\hbar l.$ Now let $L^{2}(S^{2};V_{l})$ denote the space of
square-integrable functions on $S^{2}$ with values in $V_{l}.$ Define angular
momentum operators $\hat{J}_{j}$ on this space by%
\begin{equation}
\hat{J}_{j}=L_{j}+\sigma_{j} \label{jlSigma}%
\end{equation}
where the $L_{j}$'s are the usual orbital angular momentum operators given by%
\[
L_{1}=-i\hbar\left(  x_{2}\frac{\partial}{\partial x_{3}}-x_{3}\frac{\partial
}{\partial x_{2}}\right)
\]
and relations obtained from this by cyclic permutations of the indices.

For any integer or half-integer $l$ (positive or negative), the
Schr\"{o}dinger \textquotedblleft realization\textquotedblright\ of the
associated representation of $\mathrm{\tilde{E}}(3)$ corresponding to that
value of $l$ will be a certain subspace of the Hilbert space $L^{2}%
(S^{2};V_{\left\vert l\right\vert }).$ (Here $l$ is the parameter in
(\ref{jDotxQ}) and (\ref{bl1}) in our analysis of the representations of
$\mathrm{\tilde{E}}(3).$)

\begin{definition}
\label{schr.def}For any integer or half-integer $l,$ the
\textbf{Schr\"{o}dinger Hilbert space}, denoted $\Gamma^{2}(S^{2};l),$ is the
subspace of $L^{2}(S^{2};V_{\left\vert l\right\vert })$ consisting of
functions $\psi:S^{2}\rightarrow V_{l}$ with the property that for all
$\mathbf{x}\in S^{2}$,%
\begin{equation}
\left(  \mathbf{\sigma}\cdot\mathbf{x}\right)  \psi(\mathbf{x})=r\hbar
l\psi(\mathbf{x}). \label{section}%
\end{equation}
Here, $\mathbf{\sigma}$ is defined by (\ref{sigmaj}). The norm of such a
function $\psi$ is computed as%
\[
\left\Vert \psi\right\Vert _{l}^{2}=\int_{S^{2}}\left\vert \psi(\mathbf{x}%
)\right\vert _{l}^{2}~d\mathbf{x},
\]
where $\left\vert \cdot\right\vert _{l}$ is the $\mathrm{SU}(2)$-invariant
norm on $V_{l}$ and $d\mathbf{x}$ is the surface-area measure on $S^{2}.$
\end{definition}

The notation $\Gamma$ is commonly used to denote sections of a vector bundle
over some manifold. The notation $\Gamma^{2}(S^{2};l)$ then denotes the space
of square-integrable sections of the complex line bundle over $S^{2}$ labeled
by $l.$

At each point $\mathbf{x}\in S^{2}$, the space of possible values for
$\psi(\mathbf{x})$ is one dimensional. If, for example, we take $\mathbf{x=n}%
,$ then $\psi(\mathbf{n})$ must lie in the eigenspace for $\sigma_{3}$ with
eigenvalue $\hbar l.$

If $l=1,$ the matrices $F_{j}:=\left.  d\Pi_{l}(e^{tE_{j}})/dt\right\vert
_{t=0}$ form the standard basis for $\mathrm{so}(3),$ and we may calculate
that%
\[
\mathbf{x}\cdot\mathbf{\sigma}=i\hbar\left(
\begin{array}
[c]{rrr}%
0 & -x_{3} & x_{2}\\
x_{3} & 0 & -x_{1}\\
-x_{2} & x_{1} & 0
\end{array}
\right)
\]
and thus that%
\[
(\mathbf{x}\cdot\mathbf{\sigma})(\mathbf{v})=i\hbar~\mathbf{x}\times
\mathbf{v}.
\]
The Schr\"{o}dinger Hilbert space may then be described as the space of
square-integrable functions $\mathbf{\psi}:S^{2}\rightarrow\mathbb{C}^{3}$
such that%
\begin{equation}
\frac{\mathbf{x}}{r}\times\mathbf{\psi}(\mathbf{x})=i\mathbf{\psi}%
(\mathbf{x}). \label{l1space}%
\end{equation}

We have described the Schr\"{o}dinger Hilbert space as a Hilbert space; it
remains to describe the action of the Euclidean Lie algebra on it. The action
of the position operators is simple enough: we put $X_{j}\psi(\mathbf{x})$
equal to $x_{j}\psi(\mathbf{x}).$ For the action of the angular momentum
operators, we wish to continue to use the formula in (\ref{jlSigma}). For this
to make sense, we must show that the space $\Gamma^{2}(S^{2};l)$ of functions
satisfying (\ref{section}) is invariant under the operators $\hat{J}_{j}$ in
(\ref{jlSigma}). To verify this invariance, we may easily verify that
$\Gamma^{2}(S^{2};l)$ is invariant under the action $\mathrm{SU}(2)$ given by%
\begin{equation}
(U\cdot\psi)(\mathbf{x})=\Pi_{l}(U)\psi(R_{U}\mathbf{x}), \label{suAction}%
\end{equation}
where as in Definition \ref{schr.def}, $R_{U}$ is the element of
$\mathrm{SO}(3)$ corresponding to the element $U$ of $\mathrm{SU}(2).$ The
operators $\hat{J}_{j}$ are obtained by differentiating the action of
$\exp(tF_{j})$ at $t=0.$ Since $\Gamma^{2}(S^{2};l)$ is invariant under the
group action in (\ref{suAction}), it is also invariant under the associated
Lie algebra action in (\ref{jlSigma}).

\subsection{The Segal--Bargmann Hilbert space}

We now define the Segal--Bargmann space over $S_{\mathbb{C}}^{2}$ associated
to a given value of $l$ and an arbitrary positive, matrix-valued density $\nu$
on $S_{\mathbb{C}}^{2}.$

\begin{definition}
For any integer or half-integer $l,$ the \textbf{space of holomorphic
sections} over $S_{\mathbb{C}}^{2},$ denoted $\mathcal{H}(S_{\mathbb{C}}%
^{2};l),$ is the space of holomorphic functions $\Psi:S_{\mathbb{C}}%
^{2}\rightarrow V_{\left\vert l\right\vert }$ with the property that%
\begin{equation}
(\mathbf{\sigma}\cdot\mathbf{a})(\Psi(\mathbf{a}))=r\hbar l\Psi(\mathbf{a})
\label{sectionC}%
\end{equation}
for all $\mathbf{a}\in S_{\mathbb{C}}^{2}.$ Let $\nu:S_{\mathbb{C}}%
^{2}\rightarrow\mathrm{Pos}(V_{l})$ be a continuous map into the space of
positive, self-adjoint operators on $V_{\left\vert l\right\vert }.$ Then the
\textbf{Segal--Bargmann space}, denoted $\mathcal{H}^{2}(S_{\mathbb{C}}%
^{2};l,\nu),$ associated to $\nu$ is the space of $\psi$ in $\mathcal{H}%
(S_{\mathbb{C}}^{2};l)$ such that%
\[
\left\Vert \psi\right\Vert _{\nu}^{2}:=\int_{S_{\mathbb{C}}^{2}}\left\langle
\psi(z)\left\vert \nu(z)\right\vert \psi(z)\right\rangle _{l}~dz<\infty.
\]
Here $dz$ is the $\mathrm{SO}(3;\mathbb{C})$-invariant measure on
$S_{\mathbb{C}}^{2}$ described in Section 6 in \cite{HM1}, and $\left\langle
\cdot|\cdot\right\rangle _{l}$ is the $\mathrm{SU}(2)$-invariant inner product
on $V_{l}.$
\end{definition}

We now wish to describe a map from $\Gamma^{2}(S^{2};l)$ to $\mathcal{H}%
(S_{\mathbb{C}}^{2};l).$ This map will consist of applying a smoothing
operator to $\psi\in\Gamma^{2}(S^{2};l)$ and then analytically continuing from
$S^{2}$ to $S_{\mathbb{C}}^{2}.$

\begin{proposition}
For all $\psi\in\Gamma^{2}(S^{2};l),$ consider the section $\Psi$ given by%
\begin{equation}
\Psi=\exp\left\{  -\frac{\hat{J}^{2}}{2m\alpha r^{2}\hbar}\right\}  \psi.
\label{PsiForm}%
\end{equation}
Then $\Psi$ admits a unique extension from $S^{2}$ to a holomorphic map of
$S_{\mathbb{C}}^{2}$ into $V_{l},$ and this extension is an element of
$\mathcal{H}(S_{\mathbb{C}}^{2};l).$ We refer to $\Psi$ as the
\textbf{Segal--Bargmann transform} of $\psi.$
\end{proposition}

Note the similarity between the definition (\ref{PsiForm}) of the
Segal--Bargmann transform and the formula for the coherent states in Theorem
\ref{cs.thm}. This similarity indicates that the Segal--Bargmann transform
$\Psi$ at a point $\mathbf{a}$ is simply the inner product of $\psi$ with the
coherent state $\chi_{\mathbf{a}}.$ The Segal--Bargmann transform is the more
convenient description in this case, simply because there is no continuous way
of parameterizing the coherent states, due to the nontriviality of the bundle
we are working with.

The Segal--Bargmann space and transform should be compared to the spaces in
\cite{Ba1} and \cite{Se3} in the $\mathbb{C}^{n}$ case. (See \cite{Fo} for
more information.) The proof of this result is deferred to Section
\ref{reduction.sec}, where it will be proved by reduction to the group case.
We will also see that there is a certain natural choice for $\nu$ such that
the Segal--Bargmann transform is unitary.

\subsection{Unitarity of the Segal--Bargmann transform}

For convenience of computation, let us write the operator occurring in the
exponent in (\ref{PsiForm}) as%
\begin{equation}
\frac{\hat{J}^{2}}{2m\alpha r^{2}\hbar}=\frac{1}{2}\tau\tilde{J}^{2},
\label{jTildeSquared}%
\end{equation}
where $\tilde{J}$ is a dimensionless version of $\hat{J}$ given by%
\[
\tilde{J}_{k}=\frac{\hat{J}_{k}}{\hbar}%
\]
and where $\tau$ is the dimensionless parameter given by%
\[
\tau=\frac{\hbar}{m\alpha r^{2}}.
\]
(The reader should not confuse $\tilde{J}$ with the dimensionless quantities
$L$ and $\hat{L}$ occurring in the formulas for the map $\mathbf{a}%
(\mathbf{x},\mathbf{p})$ and the annihilation operators. In computing $L,$ we
divide by $m\alpha r^{2}$ rather than by $\hbar.$)

In \cite{HM1}, we argued that $\tau$ controls the ratio of the spatial width
of the coherent states to the radius of the sphere. Specifically, if
$\Delta\mathbf{X}$ denotes the spatial width of a coherent state (measured in
some reasonable way), then we expect that%
\[
\frac{\Delta\mathbf{X}}{r}\approx\sqrt{\frac{\tau}{2}},
\]
at least when $\tau\ll1.$

\begin{theorem}
For each integer or half-integer $l$ and each $\tau>0,$ there exists a
function $\nu_{\tau}^{l}$ with values in positive operators on $V_{l}$ such
that the Segal--Bargmann transform is a unitary map of $\Gamma^{2}(S^{2};l)$
onto $\mathcal{H}^{2}(S_{\mathbb{C}}^{2};l,\nu_{\tau}^{l}).$
\end{theorem}

This result is proved in the following subsection. See Theorem
\ref{twistedSBT.thm} for a formula for $\nu_{\tau}^{l}.$

\subsection{Reduction to the group case\label{reduction.sec}}

In this subsection, we begin with the Segal--Bargmann transform for the
compact Lie group $\mathrm{SU}(2),$ as described in \cite{H1}. We then
\textquotedblleft twist\textquotedblright\ this transform with the space
$V_{l}$ carrying an irreducible representation of $\mathrm{SU}(2).$ Next, we
allow the resulting transform to descend from $\mathrm{SU}(2)$ to
$\mathrm{SU}(2)/D=S^{2},$ obtaining a Segal--Bargmann transform for $V_{l}%
$-valued functions on $S^{2},$ with respect to a \textquotedblleft
covariant\textquotedblright\ Laplacian that can be computed as the sum of
squares of the operators $\hat{J}_{j}$ in (\ref{jlSigma}). Finally, we
restrict the Segal--Bargmann transform for $V_{l}$-valued functions on $S^{2}
$ to the subspace of functions satisfying the condition (\ref{section}).

Recall from (\ref{fBasis}) the basis $\{E_{1},E_{2},E_{3}\}$ for
$\mathrm{su}(2),$ satisfying $[E_{j},E_{k}]=\varepsilon_{jkl}E_{l}.$ We then
take the inner product on $\mathrm{su}(2)$ for which these elements are
orthonormal. For each $j,$ we form the self-adjoint operator $\Sigma_{j}$ on
$L^{2}(\mathrm{SU}(2))$ (with respect to the Haar measure), given by%
\[
(\Sigma_{j}\phi)(x)=i\left.  \frac{d}{dt}\phi\left(  e^{-tE_{j}}x\right)
\right\vert _{t=0}.
\]
We then form the Laplacian $\Delta$ (here taken to be a positive operator)
given by%
\[
\Delta=\Sigma_{j}\Sigma_{j}.
\]
Using $\Delta,$ we form the \textit{heat operator} $e^{-\tau\Delta/2}.$

\begin{theorem}
\label{hall1.thm}Fix a positive number $\tau.$ Then for each $\phi\in
L^{2}(\mathrm{SU}(2)),$ the function $\Phi:=e^{-\tau\Delta/2}\phi$ admits a
holomorphic extension from $\mathrm{SU}(2)$ to $\mathrm{SL}(2;\mathbb{C}).$
Furthermore, there is a smooth positive density $\nu_{\tau}$ on $\mathrm{SL}%
(2;\mathbb{C})$ such that%
\begin{equation}
\left\Vert \phi\right\Vert _{L^{2}(\mathrm{SU}(2))}^{2}=\int_{\mathrm{SL}%
(2;\mathbb{C})}\left\vert \Phi(g)\right\vert ^{2}\nu_{\tau}(g)~dg.
\label{groupSBT}%
\end{equation}
Finally, if $\Phi$ is any holomorphic function on $\mathrm{SL}(2;\mathbb{C}) $
for which the integral on the right-hand side of (\ref{groupSBT}) is finite,
there is a unique $\phi\in L^{2}(\mathrm{SU}(2))$ for which $\left.
\Phi\right\vert _{\mathrm{SU}(2)}=e^{-\tau\Delta/2}\phi.$
\end{theorem}

This result is the $K=\mathrm{SU}(2)$ case of Theorem 2 of \cite{H1}. There is
a trivial extension of this theorem in which $\phi$ takes values in $V_{l}$
instead of in $\mathbb{C}.$ This extended Segal--Bargmann transform maps
$L^{2}(\mathrm{SU}(2);V_{l})$ (square-integrable, $V_{l}$-valued functions on
$\mathrm{SU}(2)$) to holomorphic, $V_{l}$-valued functions on $\mathrm{SL}%
(2;\mathbb{C}).$ To get something slightly less trivial, we are going to
\textquotedblleft twist\textquotedblright\ our functions by the action of
$\mathrm{SU}(2)$ on $V_{l}.$ When we apply the associated \textquotedblleft
twisted Laplacian\textquotedblright\ to functions that are invariant under the
right action of the diagonal subgroup $D$ of $\mathrm{SU}(2),$ we obtain
precisely the operator (\ref{jTildeSquared}). The result of applying Theorem
\ref{hall1.thm} to functions of this type is the following.

\begin{theorem}
\label{twistedSBT.thm}Let $\psi$ be any function in $L^{2}(S^{2};V_{l})$ and
let $\psi$ be the holomorphic function on $S_{\mathbb{C}}^{2}$ whose
restriction to $S^{2}$ is given by%
\[
\Psi=\exp\left\{  -\tau\tilde{J}^{2}/2\right\}  \psi.
\]
Then%
\[
\left\Vert \psi\right\Vert _{L^{2}(S^{2};V_{l})}^{2}=\int_{S_{\mathbb{C}}^{2}%
}\left\langle \Psi(z),\nu_{\tau}^{l}(z)\Psi(z)\right\rangle _{l}~dz,
\]
where $\nu_{\tau}^{l}$ is the function with values in positive operators on
$V_{l}$ given by%
\[
\nu_{\tau}^{l}(R_{g}\mathbf{n})=\int_{D_{\mathbb{C}}}\Pi_{l}\left(
((gh)^{-1})^{\ast}(gh)^{-1}\right)  \nu_{\tau}(gh)~dh
\]
for each $g\in\mathrm{SL}(2;\mathbb{C}).$ Here $R_{g}$ is the element of
$\mathrm{SO}(3;\mathbb{C})$ associated to $g\in\mathrm{SL}(2;\mathbb{C}).$
Furthermore, if $\psi$ has the property (\ref{section}), the $\Psi$ has the
property (\ref{sectionC}).
\end{theorem}

Using a slight variant of the method of Flensted-Jensen \cite{FJ}, one can
show that the function $\nu_{\tau}^{l}$ satisfies a bundle heat equation over
hyperbolic 2-space, which is the noncompact symmetric space dual (in the usual
duality between compact and noncompact symmetric spaces) to $S^{2}. $

\begin{proof}
To each function $\phi\in L^{2}(\mathrm{SU}(2);V_{l}),$ let us associate
another function $\tilde{\phi}$ given by%
\[
\tilde{\phi}(x)=\Pi_{l}(x)\phi(x).
\]
Since $\Pi_{l}$ is unitary, $\phi$ and $\tilde{\phi}$ have the same norm.

For $y\in\mathrm{SU}(2),$ let $L_{y}$ denote the \textquotedblleft
ordinary\textquotedblright\ left action of $y$ on some $\psi\in L^{2}%
(\mathrm{SU}(2);V_{l}),$ namely%
\[
(L_{y}\phi)(x)=\phi(y^{-1}x).
\]
We may also introduce the twisted left action $\tilde{L}_{y}$ given by%
\[
(\tilde{L}_{y}\psi)(x)=\Pi(y)\psi(y^{-1}x).
\]
It is easily verified that%
\[
\widetilde{\left(  L_{y}\phi\right)  }=\tilde{L}_{y}\tilde{\phi}.
\]

Differentiating this relation, we find that%
\[
\widetilde{\left(  \Sigma_{j}\phi\right)  }=\tilde{\Sigma}_{j}\tilde{\phi},
\]
where%
\[
(\tilde{\Sigma}_{j}\psi)(x)=\frac{d}{dt}\psi\left(  e^{-tE_{j}}x\right)
+\pi_{l}(E_{j})\psi(x).
\]
It is then easy to see that%
\[
\widetilde{\left(  e^{-\tau\Delta/2}\phi\right)  }=e^{-\tau\tilde{\Delta}%
/2}\tilde{\phi},
\]
where%
\[
\tilde{\Delta}=\tilde{\Sigma}_{j}\tilde{\Sigma}_{j}.
\]
Thus, if $\Phi$ is the holomorphic extension of $e^{-\tau\Delta/2}\phi$ and
$\tilde{\Phi}$ is the holomorphic extension of $e^{-\tau\tilde{\Delta}%
/2}\tilde{\phi},$ we have%
\begin{equation}
\Phi(g)=\Pi(g^{-1})\tilde{\Phi}(g). \label{psiPsi}%
\end{equation}
Let us now apply the Segal--Bargmann transform for $\mathrm{SU}(2)$ (trivially
extended to $V_{l}$-valued functions), to $\psi,$ and express the result in
terms of $\tilde{\Psi}$ by means of (\ref{psiPsi}):%
\begin{align}
\left\Vert \tilde{\phi}\right\Vert _{L^{2}(\mathrm{SU}(2);V_{l})}^{2}  &
=\left\Vert \phi\right\Vert _{L^{2}(\mathrm{SU}(2);V_{l})}^{2}\nonumber\\
&  =\int_{\mathrm{SL}(2;\mathbb{C})}\left\langle \Phi(g),\Phi(g)\right\rangle
_{l}\nu_{\tau}(g)~dg\nonumber\\
&  =\int_{\mathrm{SL}(2;\mathbb{C})}\left\langle \Pi(g^{-1})\tilde{\Phi
}(g),\Pi(g^{-1})\tilde{\Phi}(g)\right\rangle _{l}\nu_{\tau}(g)~dg\nonumber\\
&  =\int_{\mathrm{SL}(2;\mathbb{C})}\left\langle \tilde{\Phi}(g),\Pi
(g^{-1})^{\ast}\Pi(g^{-1})\tilde{\Phi}(g)\right\rangle _{l}\nu_{\tau
}(g)~dg\nonumber\\
&  =\int_{\mathrm{SL}(2;\mathbb{C})}\left\langle \tilde{\Phi}(g),\Pi\left(
(g^{-1})^{\ast}g^{-1}\right)  \tilde{\Phi}(g)\right\rangle _{l}\nu_{\tau
}(g)~dg. \label{psiTilde}%
\end{align}

Now, let $D$ be the diagonal subgroup of $\mathrm{SU}(2),$ so that
$\mathrm{SU}(2)/D=S^{2},$ and let $D_{\mathbb{C}}$ be the complexification of
$D,$ which is just the diagonal subgroup of $\mathrm{SL}(2;\mathbb{C}).$ It is
easy to see that the twisted left action of $\mathrm{SU}(2)$ commutes with the
\textit{ordinary} right action of $\mathrm{SU}(2).$ It follows that the space
of functions on $\mathrm{SU}(2)$ that are invariant under the ordinary right
action of $D$ is invariant under $\tilde{\Delta}$ and thus under the heat
operator $e^{-\tau\tilde{\Delta}/2}.$ Thus, if we apply (\ref{psiTilde}) in
the case that $\tilde{\phi}$ is invariant under the ordinary right action of
$D,$ $\tilde{\Phi}$ will be invariant under the ordinary right action of $D$
and thus also (because $\tilde{\Phi}$ is holomorphic) under the ordinary right
action of $D_{\mathbb{C}}.$ Meanwhile, we can break up the integration over
$\mathrm{SL}(2;\mathbb{C})$ into an integral over $D_{\mathbb{C}}$ followed by
an integration over $\mathrm{SL}(2;\mathbb{C})/D_{\mathbb{C}}.$ Thus,
(\ref{psiTilde}) becomes, when $\tilde{\phi}$ is right-$D$-invariant,%
\begin{align*}
&  \left\Vert \tilde{\phi}\right\Vert _{L^{2}(\mathrm{SU}(2);V_{l})}^{2}\\
&  =\int_{\mathrm{SL}(2;\mathbb{C})/D_{\mathbb{C}}}\int_{D_{\mathbb{C}}%
}\left\langle \tilde{\Phi}(g),\Pi\left(  ((gh)^{-1})^{\ast}(gh)^{-1}\right)
\tilde{\Phi}(g)\right\rangle _{l}\nu_{\tau}(gh)~dh~d[g],
\end{align*}
where $d[g]$ is the $\mathrm{SL}(2;\mathbb{C})$-invariant volume measure on
$\mathrm{SL}(2;\mathbb{C})/D_{\mathbb{C}}.$ After identifying $\mathrm{SL}%
(2;\mathbb{C})/D_{\mathbb{C}}$ with $S_{\mathbb{C}}^{2}$ and $d[g]$ with the
invariant volume measure on $S_{\mathbb{C}}^{2},$ we obtain the first claimed
result in the theorem.

If $\psi$ ($=\tilde{\phi}$) is a right-$D$-invariant function on
$\mathrm{SU}(2),$ it descends to a function on $\mathrm{SU}(2)/D=S^{2}.$
Furthermore, the action $\tilde{\Sigma}_{j}$ on this function corresponds to
the action of the $\tilde{J}_{j}$ on the associated function on the sphere.
Thus, the twisted Segal--Bargmann transform for $\psi$ is just the transform
associated to the operator $\exp(-\tau\tilde{J}^{2}/2)$ on $S^{2}.$ Meanwhile,
it is easily seen operators $\tilde{J}_{j}$ preserve the condition
(\ref{section}). Thus, we can specialize our transform on $L^{2}(S^{2};V_{l})$
to the subspace $\Gamma^{2}(S^{2};l).$
\end{proof}

\end{document}